\pdfoutput=1
\documentclass{article}
\usepackage{spconf,amsmath,graphicx}
\usepackage{tabto}
\usepackage[table]{xcolor}
\usepackage[justification=centering]{caption}
\usepackage{calrsfs}
\usepackage[ruled,vlined]{algorithm2e}
\usepackage{graphics}
\usepackage{float}
\DeclareMathAlphabet{\pazocal}{OMS}{zplm}{m}{n}
\usepackage{mathtools} 
\usepackage{amssymb}
\usepackage{booktabs}
\usepackage{siunitx}
\usepackage{amsmath}
\usepackage{multirow}
\usepackage{booktabs,chemformula}
\newcommand{\Lb}{\pazocal{L}}

\title{LEVERAGING 3D INFORMATION IN UNSUPERVISED BRAIN MRI SEGMENTATION}
\name{\begin{tabular}{c}
    Benjamin Lambert$^1$, \qquad Maxime Louis$^{1}$, \qquad Senan Doyle$^1$, \qquad Florence Forbes$^2$  \\
    Michel Dojat$^{3}$, \qquad Alan Tucholka$^1$, \qquad \normalfont{for the Alzheimer’s Disease Neuroimaging Initiative}* 
\end{tabular}}

\address{
$^1$ Pixyl, Research and Development Laboratory, 38000 Grenoble, France \\
$^2$ Univ. Grenoble Alpes, Inria, CNRS, Grenoble INP, LJK, 38000 Grenoble, France \\
$^3$ Univ. Grenoble Alpes, Inserm, U1216, Grenoble Institut Neurosciences, GIN, 38000 Grenoble, France}

\newcommand\blfootnote[1]{
  \begingroup
  \renewcommand\thefootnote{}\footnote{#1}
  \addtocounter{footnote}{-1}
  \endgroup
}

\usepackage{hyperref}
\begin{document}

\maketitle

\begin{abstract}
Automatic segmentation of brain abnormalities is challenging, as they vary considerably from one pathology to another. Current methods are supervised and require numerous annotated images for each pathology, a strenuous task. To tackle anatomical variability, Unsupervised Anomaly Detection (UAD) methods are proposed, detecting anomalies as outliers of a healthy model learned using a Variational Autoencoder (VAE). Previous work on UAD adopted a 2D approach, meaning that MRIs are processed as a collection of independent slices. Yet, it does not fully exploit the spatial information contained in MRI. Here, we propose to perform UAD in a 3D fashion and compare 2D and 3D VAEs. As a side contribution, we present a new loss function guarantying a robust training. Learning is performed using a multicentric dataset of healthy brain MRIs, and segmentation performances are estimated on White-Matter Hyperintensities and tumors lesions. Experiments demonstrate the interest of 3D methods which outperform their 2D counterparts.

\blfootnote{*Data used in preparation of this article were partially obtained from the Alzheimer’s Disease
Neuroimaging Initiative (ADNI) database (adni.loni.usc.edu). As such, the investigators
within the ADNI contributed to the design and implementation of ADNI and/or provided data
but did not participate in analysis or writing of this report. A complete listing of ADNI
investigators can be found at: \url{http://adni.loni.usc.edu/wp-content/uploads/how_to_apply/ADNI_Acknowledgement_List.pdf}}

\end{abstract}
\vspace{-1mm}
\section{Introduction}
\label{sec:intro}
Brain anomalies are widely used as biomarkers indicating the presence or progress of many neurological disorders. Magnetic Resonance Imaging (MRI) is today an essential modality to reveal these markers. In recent years, a multitude of algorithms for the automatic detection of anomalies in brain MRI has been proposed, with promising results achieved by Deep Learning (DL) approaches.

Anomalies can manifest in a wide range of shape, location and intensities. To automatically detect them, most state-of-the-art algorithms use a supervised approach, trained on an as large as possible manually annotated dataset \cite{akkus2017deep}. Acquisition of such dataset can be both time-consuming and expensive. Besides, these algorithms are specific to the task they are trained for and will therefore perform poorly when applied on images presenting unseen type of anomalies.

To overcome these limitations, Unsupervised Anomaly Detection (UAD) methods are proposed. They consist in the construction of a healthy model, which allows to detect anomalies from their deviation from the model. Recent DL methods propose to learn a manifold of healthy patients by computing lower-dimensional latent representations of the scans.  At inference, anomalous images are mapped to the learned healthy manifold. An anomaly map is then computed using the voxel-wise difference from its healthy projection. Although failing for the time being to meet the state-of-the-art results of supervised methods, UAD approaches have several advantages: they do not require any labeled data and can be adapted with very limited adjustments to different types of pathologies. 

Two types of neural networks are particularly investigated in UAD : Generative Adversarial Networks (GANs) \cite{goodfellow2014generative} and Variational Autoencoders (VAEs) \cite{kingma2013auto}. GANs use an adversarial training approach to construct the healthy manifold \cite{di2019survey}. VAEs rather encode images into latent distributions and use a regularization term in their loss function to ensure consistency in the latent space \cite{baur_autoencoders_2020}. Yet, training  of VAEs is known to be unstable \cite{lucas_understanding_2019}. As an attempt to alleviate this issue, we propose a robust loss function to ensure a stable training.

To the best of our knowledge, all previous research related to UAD tackle the problem in a 2D fashion, meaning that the MRI volume is processed as a collection of independent 2D slices. Despite being computationally efficient, this approach has several limitations. First, it requires the selection of homogeneous slices. Top and bottom slices, which contain no or little brain tissue, are excluded, as they may compromise the training of 2D networks.  It is an impairing step because it requires hand-tuning and prevents the detection of anomalies in the entire MRI. Second, the processing of 2D data does not fully exploit the spatial information available in the scans. This restriction to 2D may arise from the significant difficulty of 3D training, which suffers from the high dimensionality of data and the reduction of the number of training samples. Another limitation of the state of the art is the restriction of most previous work on UAD to monocentric datasets for training \cite{baur_autoencoders_2020, chen2018unsupervised, zimmerer_context-encoding_2020}, which fails to ensure a clinically realistic setup. 

In this work, we present a 3D VAE framework for UAD. We train 2D and 3D models using our proposed robust loss function and compare them on multiple pathologies acquired in real clinical conditions. Experiments show that exploiting the full images in a 3D fashion leads to better performances than the usual 2D approaches. 

\section{Methods}
\label{sec:methods}

\subsection{UAD using Variational Autoencoders} 
 \label{ssec:VAE_UAD}
VAEs are composed of an encoder followed by a decoder. The encoder compresses the input image $X$ into a distribution over the latent space, which is regularized to be close to a prior distribution conventionally set to be a Gaussian multivariate distribution $N(z;0,I)$. A point $z$ is then sampled from this latent distribution and presented to the decoder which produces a reconstruction $\hat{X}$ of the input. 
 
 The bottleneck separating the encoder and the decoder distinguishes spatial and dense VAEs. In the spatial configuration, VAEs are fully-convolutional, meaning that the latent vector is a multi-dimensional tensor $z\in\mathbb{R}^{N \times h \times w}$ in 2D and $z\in\mathbb{R}^{N\times d \times h \times w}$ in 3D with $N$ the latent space dimension, $h$ the height, $w$ the width and $d$ the depth. In this configuration, spatial information is preserved during the encoding-decoding process. In the dense configuration, the image is encoded into a 1-dimensional tensor $z\in\mathbb{R}^N$ with the use of fully-connected layers, and as a consequence spatial information is discarded.

After training on healthy subjects, we perform UAD using a reconstruction-based approach. Reconstruction is the classical approach for DL-based UAD. It consists in the computation of anomaly maps by subtracting its reconstruction $\hat{X}$ to the input image $X$. As a result of the training, anomalies are badly reconstructed once projected to the healthy manifold, and thus yield to high values in the anomaly map. By binarizing them with an appropriate threshold, anomalies segmentations are obtained.

\subsection{Collapsing-robust loss function} \label{sssec:Collapsing-robust loss function}

Training of a VAE is performed by minimizing a two-terms loss function: the evidence lower-bound ($ELBO(X,\hat{X})$): 
\begin{equation}
    \Lb  = \vert X-\hat{X} \vert_d + D_{KL}(q_\phi(z|X)|P(z)), d \in \{1,2\}
\label{eqn:VAE_loss}
\end{equation}

The left-hand side term of this equation is the reconstruction term, which computes the distance between the image $X$ and its reconstruction $\hat{X}$. The second term is the Kullback-Leibler Divergence (KL) between the posterior $q_\phi(z|X)$ and the defined prior $P(z)$, which acts as a regularization term in the latent space. 
Training of VAE using equation \eqref{eqn:VAE_loss} is prone to \emph{posterior collapse}, meaning that the network may learn to neglect a subset of latent variables to match the prior, and thus the generative power of the network degrades \cite{lucas_understanding_2019}. Solutions have been proposed to tackle the problem, including the addition of a hyperparameter $\beta$ to balance both terms of the equation \cite{higgins2016beta}. Following the same motivation, KL annealing schedules have been presented, including the KL cyclical annealing schedule described in \cite{fu2019cyclical}. We build on this work and propose a custom loss function defined as follows:
\begin{align}
    \Lb_{T} &=  \dfrac{\vert X-\hat{X} \vert_1}{\Sigma} + \beta(t) D_{KL}(q_\phi(z|X)|P(z)) \label{eq:Collapsing_robust_loss1} \\ 
    \text{with } \beta(t) &=
    \left\{
    \begin{array}{ll}
        \dfrac{2t}{T} \text{ for t} \in [0, \dfrac{T}{2}[ \nonumber \\  
        \hspace{2pt} 1 \hspace{8pt} \text{else} \nonumber     \end{array}
\right.
\text{and } \beta(t+T)=\beta(t)
\end{align}
where $\Sigma$ is the moving mean computed on the $L$ last values of the reconstruction term, $t$ is the current iteration and T the cycle duration. $T$ and $L$ are two hyperparameters that we set to 50 and 10 respectively in our experiments. We choose the $\ell_1$ norm as reconstruction term to reduce blurring of the reconstructed images. Normalizing this term by its moving mean maintains the reconstruction term near one, which encourages the network to learn throughout the entire training stage even when reconstruction becomes satisfying. Coupled with the KL cyclical annealing schedule, we obtain a loss robust to collapsing which ensures a stable training of VAEs. We use this loss without any additional parameter tuning in all our experiments.

\subsection{Architecture Details}  \label{sssec:Architectures} 

Our encoders are composed of 6 layers: 4 convolutional layers compressing the input image, and 2 additional layers to encode the result into a normal distribution by computing its parameters $\mu$ and $\sigma$. These two last layers are convolutional in the spatial configuration, and fully-connected for dense VAEs. A latent vector $z$ is then sampled from this distribution and passed to the decoder. 

Decoders adopt a symmetric architecture with 5 layers decompressing $z$ in order to reconstruct the input image.  Similarly, the first layer of the decoder is convolutional in the spatial configuration, and fully-connected otherwise. 3D adaptations of our architectures are obtained by replacing 2D convolutions by their 3D counterparts. 



\section{Experiments}
\label{sec:experiments}


We focus on brain MRI FLAIR. To train our networks, we gather 79 healthy scans from several opensource datasets : the ADNI dataset, IBC \footnotemark \cite{individual_brain_chart} and Kirby \cite{kirby}. Evaluation of our networks is performed on 2 different datasets presenting various brain lesions. We use a collection of 196 scans with annotated White-Matter Hyperintensities (WMH) gathered from several opensource datasets : the MICCAI MSSEG Challenge \cite{msseg_dataset}, the ISBI MS Lesion Challenge \cite{ISBI_dataset}, and the WMH Challenge \cite{WMH_dataset}. Additionally, we select 100 brain tumors scans from the BraTS 2018 dataset \cite{brats1}. These datasets are multi-centric and comprise various scanners and clinical protocols, hence assuring a clinically realistic setup. 

\footnotetext{This data was obtained from the OpenfMRI database. Its accession number is ds000244.}

 Scans are rigidly registered on a $170\times204\times170$ template with an isotropic resolution of $1mm$, corrected for bias using \cite{tustison2010n4itk} and skull-stripped using the HD-BET algorithm \cite{isensee2019automated}. Volumes are then cropped to $160\times192\times75$ to concentrate on central slices, for which brain tissue is abundant. Finally, intensity range is set to $[0,1]$. 

Segmentations are produced by binarizing anomaly maps, obtained as the difference between an image and its reconstruction. Binarization is performed using a threshold above which voxels are considered as anomalies. In order to find its optimal value, we split each testing set in half. On the first half, we compute Dice scores between ground truths and segmentations obtained with 15 different thresholds, ranged between $0 \text{ and } 0.15$, which we found to be a relevant range in our experiments. Then, the threshold which provided the best performance is applied on the second half. 

 The obtained segmentations are then multiplied by a slightly eroded brain mask, in order to remove false positives occurring near the brain contour. Finally, a median filter is applied to remove regions with less than 10 voxels.
 


\begin{figure*}[!ht]
\centering
\includegraphics[width=17cm]{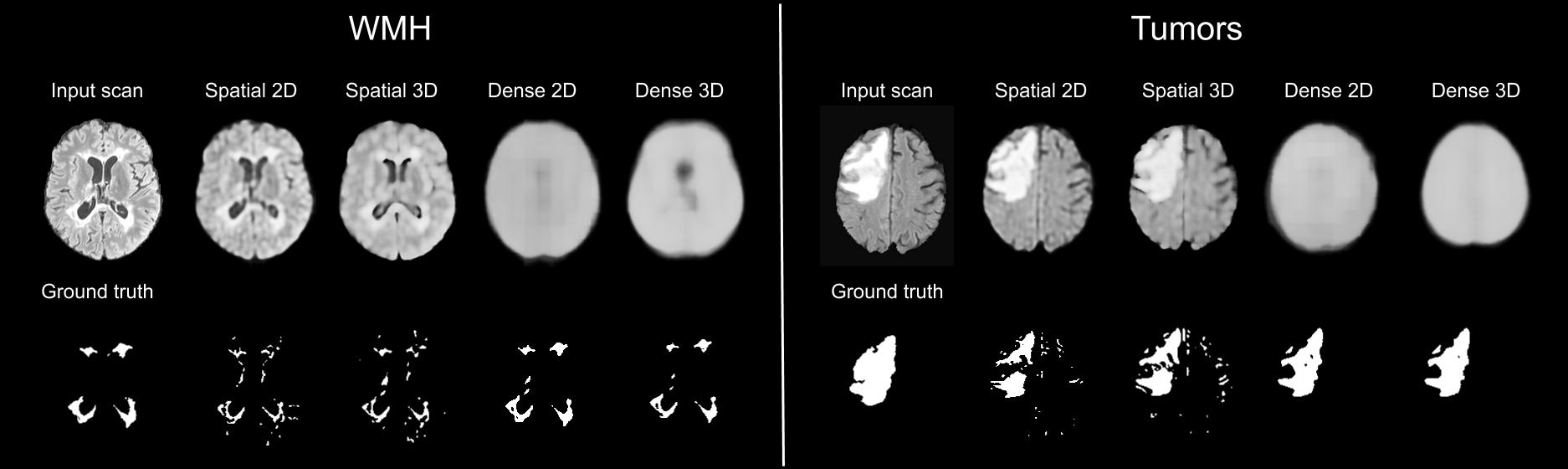}
\caption{Reconstructions and segmentations of our networks. Left : WMH lesions. Right : tumors lesions.}
\label{fig:results}
\end{figure*}

\section{Results} \label{sec:Results}

We evaluate our networks segmentation performance using Dice,  sensitivity and specificity scores. Reconstruction performance is estimated using the voxel-wise mean average error (MAE) between the input scan and its reconstruction. To fairly compare 2D and 3D networks, they are all evaluated on the same 75 central slices for each test scan. Performances of the Spatial 2D and Dense 2D, as well as their 3D adaptations, are presented in Table \ref{tab:Results}. Illustrative reconstructions and segmentations can be found in Figure \ref{fig:results} for both WMH and tumor lesions. 3D adaptations demonstrate a boost in performance as compared to 2D. On both the WMH and Tumors datasets, the best network is the Dense 3D VAE, with dice scores of $0.463\pm0.259$ and $0.650\pm0.190$, a specificity of $0.996$ and $0.981$, and a sensitivity of $0.511$ and $0.711$ respectively.

\begin{table*}[!ht]
\centering
\begin{tabular}{|c|c|c|c|c|c|c|c|c|}
    \hline
     & \multicolumn{4}{c|}{WMH} & \multicolumn{4}{c|}{Tumors} \\
      \hline
     Model & Dice & Spe & Sen & MAE & Dice & Spe & Sen & MAE  \\
    \hline
    Spatial 2D  & $0.280\pm0.174$ & $0.986$ & $0.547$ & $0.049$ & $0.260\pm0.121$ & $0.912$ & $0.376$ & $0.079$ \\
    \hline
    Spatial 3D  & $0.336\pm0.203$ & $0.989$ & $0.605$ & $0.056$ & $0.386\pm0.160$ & $0.898$ & $0.655$ & $0.065$  \\
    \hline
    Dense 2D & $0.460\pm0.262$ & $0.994$ & $0.604$ & $0.102$ & $0.618\pm0.167$ & $0.978$ & $0.676$ & $0.133$  \\
    \hline
    Dense 3D  & $\textbf{0.463}\pm\textbf{0.259}$ & $0.996$ & $0.511$ & $0.104$ & $\textbf{0.650}\pm\textbf{0.190}$ & $0.981$ & $0.711$ & $0.151$  \\
    \hline
\end{tabular}
\caption{\label{tab:Results}
Performance metrics obtained on the test datasets. Spe = Specificity, Sen = Sensitivity, MAE = Mean Average Error.}
\end{table*}


\section{Discussion and Conclusion} \label{sec:Conclusion}
In this work, the interest of 3D methods for UAD in brain MRIs was evaluated. Using our proposed collapsing-robust loss function, 2D and 3D models were compared on two pathological datasets showing an overall increase in performance in the 3D setting. For spatial VAEs, this gain was more significant than for dense networks. In the dense configuration, spatial information was not preserved during the encoding-decoding scheme. Thus, the introduction of additional spatial context with the use of 3D did not significantly increase the segmentation performance. Additionally, we observed that dense networks outperformed spatial networks in both the 2D and 3D cases. They indeed operated a significantly higher compression of the scan as compared to spatial networks. Consequently, reconstructions presented very little details, and anomalies were entirely discarded, which made them more easily detectable. 

 Performances of our approaches, although being satisfying, remain lower than state-of-the-art supervised methods. In \cite{lachinov2018glioma}, a Dice score above $0.9$ is obtained in tumor segmentation task, and a score of $0.56$ is achieved in \cite{valverde2017improving} for WMH lesions segmentation with supervised approaches. Although UAD methods fail at reaching these standards, they are very promising as they allow to detect anomalies in a generic fashion and without the use of annotated training samples. One interesting application of UAD approaches is the generation of raw segmentations, which can be used as a starting point for manual segmentations, saving time for the rater. 

 In future work, advanced 3D architectures could be tested with the use of deeper networks. Extensions to multi-sequential MRI data is also of interest.

\section{Compliance with ethical standards}
This research study was conducted retrospectively using
human subject data made available by the following sources: ADNI, OpenfMRI, NITRC, the MICCAI MSSEG Challenge, the ISBI MS Lesion Challenge, the WMH Challenge and BraTS 2018. Ethical approval was not required as confirmed by the license attached with the data. 
\section{Acknowledgments}
BL, ML, SD and AT are employees of the Pixyl Company. MD and FF serve on Pixyl advisory board.

\bibliographystyle{IEEEbib}
\bibliography{reference}

\begin{thebibliography}{10}

\bibitem{akkus2017deep}
Z.~Akkus et~al.,
\newblock ``Deep learning for brain {{MRI}} segmentation: state of the art and
  future directions,''
\newblock {\em J Dig Im}, vol. 30, no. 4, 2017.

\bibitem{goodfellow2014generative}
Ian Goodfellow et~al.,
\newblock ``Generative adversarial nets,''
\newblock {\em Adv Neural Inf Proc Sys}, 2014.

\bibitem{kingma2013auto}
D.~P. Kingma et~al.,
\newblock ``Auto-encoding variational {Bayes},''
\newblock {\em Int Conf Learn Repres}, 2013.

\bibitem{di2019survey}
F.~Di~Mattia et~al.,
\newblock ``A survey on {{GAN}}s for anomaly detection,''
\newblock {\em CoRR}, 2019.

\bibitem{baur_autoencoders_2020}
C.~Baur et~al.,
\newblock ``Autoencoders for unsupervised anomaly segmentation in brain {{MR}}
  images: A comparative study,''
\newblock {\em CoRR}, 2020.

\bibitem{lucas_understanding_2019}
J.~Lucas et~al.,
\newblock ``Understanding posterior collapse in generative latent variable
  models,''
\newblock {\em Int Conf Learn Repres}, 2019.

\bibitem{chen2018unsupervised}
X.~Chen et~al.,
\newblock ``Unsupervised detection of lesions in brain {{MRI}} using
  constrained adversarial auto-encoders,''
\newblock {\em CoRR}, 2018.

\bibitem{zimmerer_context-encoding_2020}
D.~Zimmerer et~al.,
\newblock ``Context-encoding variational autoencoder for unsupervised anomaly
  detection,''
\newblock {\em CoRR}, 2018.

\bibitem{higgins2016beta}
I.~Higgins et~al.,
\newblock ``beta-{{VAE}}: Learning basic visual concepts with a constrained
  variational framework,''
\newblock {\em Int Conf Learn Repres}, 2016.

\bibitem{fu2019cyclical}
H.~Fu et~al.,
\newblock ``Cyclical annealing schedule: A simple approach to mitigating {{KL}}
  vanishing,''
\newblock {\em NAACL-HLT}, vol. 1, 2019.

\bibitem{individual_brain_chart}
A.~L. Pinho et~al.,
\newblock ``Individual brain charting, a high-resolution fmri dataset for
  cognitive mapping,''
\newblock {\em Scientific data}, vol. 5, 2018.

\bibitem{kirby}
B.~A. Landman et~al.,
\newblock ``Multi-parametric neuroimaging reproducibility: a {3T} resource
  study,''
\newblock {\em NeuroIm}, 2011.

\bibitem{msseg_dataset}
Olivier Commowick et~al.,
\newblock ``Objective evaluation of multiple sclerosis lesion segmentation
  using a data management and processing infrastructure,''
\newblock {\em Scientific reports}, vol. 8, no. 1, 2018.

\bibitem{ISBI_dataset}
Aaron Carass et~al.,
\newblock ``Longitudinal multiple sclerosis lesion segmentation: resource and
  challenge,''
\newblock {\em NeuroImage}, vol. 148, 2017.

\bibitem{WMH_dataset}
Hugo~J. Kuijf et~al.,
\newblock ``Standardized assessment of automatic segmentation of white matter
  hyperintensities and results of the wmh segmentation challenge,''
\newblock {\em IEEE transactions on medical imaging}, vol. 38, no. 11, 2019.

\bibitem{brats1}
B.~H. Menze et~al.,
\newblock ``The multimodal brain tumor image segmentation benchmark
  ({{BRATS}}),''
\newblock {\em TMI}, vol. 34, no. 10, 2014.

\bibitem{tustison2010n4itk}
N.~J. Tustison et~al.,
\newblock ``N4itk: improved {{N3}} bias correction,''
\newblock {\em TMI}, vol. 29, no. 6, 2010.

\bibitem{isensee2019automated}
F.~Isensee et~al.,
\newblock ``Automated brain extraction of multisequence {{MRI}} using
  artificial neural networks,''
\newblock {\em HBM}, vol. 40, no. 17, 2019.

\bibitem{lachinov2018glioma}
D.~Lachinov et~al.,
\newblock ``Glioma segmentation with cascaded {{UN}}et,''
\newblock in {\em BrainLes}. Springer, 2018.

\bibitem{valverde2017improving}
S.~Valverde et~al.,
\newblock ``Improving automated multiple sclerosis lesion segmentation with a
  cascaded {{3D}} convolutional neural network approach,''
\newblock {\em NeuroIm}, vol. 155, 2017.

\end{thebibliography}

\end{document}